\documentclass[11pt]{article}
\pdfoutput=1
\usepackage{fullpage}
\usepackage{xcolor}
\usepackage{cite}
\usepackage{graphicx}
\usepackage{tensor}
\usepackage{amsmath}
\usepackage{amssymb}
\usepackage{subcaption}
\usepackage[utf8x]{inputenc} 
\title{}
\date{}
\renewcommand{\vec}[1]{\mbox{\boldmath$ #1 $}}

\def\beq{\begin{equation}}
\def\eeq{\end{equation}}
\allowdisplaybreaks

\renewcommand{\Re}{\operatorname{Re}}

\begin{document}
\bibliographystyle{utphys}

\newcommand{\hel}{\eta} 
\renewcommand{\d}{\mathrm{d}}
\newcommand{\dd}{\hat{\mathrm{d}}}
\newcommand{\del}{\hat{\delta}}
\newcommand{\ket}[1]{| #1 \rangle}
\newcommand{\bra}[1]{\langle #1 |}

\newcommand{\be}{\begin{equation}}
\newcommand{\ee}{\end{equation}}
\newcommand\n[1]{\textcolor{red}{(#1)}} 
\newcommand{\diff}{\mathop{}\!\mathrm{d}}
\newcommand{\lb}{\left}
\newcommand{\rb}{\right}
\newcommand{\f}{\frac}
\newcommand{\pd}{\partial}
\newcommand{\tr}{\text{tr}}
\newcommand{\fdiff}{\mathcal{D}}
\newcommand{\im}{\text{im}}
\let\caron\v
\renewcommand{\v}{\mathbf}
\newcommand{\T}{\tensor}
\newcommand{\R}{\mathbb{R}}
\newcommand{\C}{\mathbb{C}}
\newcommand{\Z}{\mathbb{Z}}
\newcommand{\msbar}{\ensuremath{\overline{\text{MS}}}}
\newcommand{\DIS}{\ensuremath{\text{DIS}}}
\newcommand{\abar}{\ensuremath{\bar{\alpha}_S}}
\newcommand{\bb}{\ensuremath{\bar{\beta}_0}}
\newcommand{\rc}{\ensuremath{r_{\text{cut}}}}
\newcommand{\Nd}{\ensuremath{N_{\text{d.o.f.}}}}
\newcommand{\red}[1]{{\color{red} #1}}

\newcommand{\Ad}{\dot{A}}
\newcommand{\Bd}{\dot{B}}
\newcommand{\Cd}{\dot{C}}
\newcommand{\Dd}{\dot{D}}
\newcommand{\Ed}{\dot{E}}
\newcommand{\Fd}{\dot{F}}
\newcommand{\depsilon}{\epsilon}
\newcommand{\dsigma}{\bar{\sigma}}

\newcommand{\bphi}{\phi} 
\newcommand{\bB}{B} 
\newcommand{\bH}{H} 
\newcommand{\bsigma}{\sigma} 
\newcommand{\charge}{\tilde{c}} 
\newcommand{\ampA}{\mathcal{A}} 
\newcommand{\ampM}{\mathcal{M}} 

\titlepage

\vspace*{0.5cm}

\begin{center}
{\bf \Large A spinorial double copy for ${\cal N}=0$ supergravity}

\vspace*{1cm} 
\textsc{Kymani Armstrong-Williams\footnote{k.t.k.armstrong-williams@qmul.ac.uk} and
Chris D. White\footnote{christopher.white@qmul.ac.uk}} \\

\vspace*{0.5cm} Centre for Theoretical Physics, Department of
Physics and Astronomy, \\
Queen Mary University of London, 327 Mile End
Road, London E1 4NS, UK\\

\end{center}

\vspace*{0.5cm}

\begin{abstract}
  The Weyl double copy is a formula relating solutions of scalar,
  gauge and gravity theories, and is related to the BCJ double copy
  for scattering amplitudes. The latter relates Yang-Mills theory to
  ${\cal N}=0$ supergravity, where an axion and dilaton are present in
  addition to the graviton. However, the traditional Weyl double copy
  applies only to pure gravity solutions, such that it remains to be
  seen whether or not it can be extended to the full spectrum of
  ${\cal N}=0$ supergravity. We examine this question using recently
  developed twistor methods, showing that some sort of double copy
  formula for ${\cal N}=0$ supergravity is indeed possible for certain
  solutions. However, it differs both from the traditional Weyl double
  copy form, and recent conjectures aimed at generalising the Weyl
  double copy to non-vacuum solutions.
\end{abstract}

\vspace*{0.5cm}

\section{Introduction}
\label{sec:intro}

Within the last couple of decades, a duality known as the {\it double
  copy} has generated a good deal of interest \cite{Bern:2019prr}. It
was first observed in the 1980s as a relationship between tree-level
scattering amplitudes for open and closed strings~\cite{Kawai:1985xq},
known as the {\it KLT relations}. In taking the low energy limit, open
and closed strings give rise to (non-abelian) gauge bosons and
gravitons respectively, such that a relationship -- the double
copy~\cite{Bern:2010ue,Bern:2010yg} -- should hold between field
theory scattering amplitudes. Importantly, the field theory double
copy has been extended to loop level, going beyond its original
stringy context, thus providing tantalising glimpses of deep common
structures underlying our various theories of fundamental
interactions. Further work has broadened the double copy beyond
scattering amplitudes at fixed order in the coupling
constant~\cite{Oxburgh:2012zr,Vera:2012ds,Johansson:2013nsa,Saotome:2012vy},
and also to classical solutions. The latter may be
exact~\cite{Monteiro:2014cda,Luna:2015paa,Ridgway:2015fdl,Bahjat-Abbas:2017htu,Carrillo-Gonzalez:2017iyj,CarrilloGonzalez:2019gof,Bah:2019sda,Alkac:2021seh,Alkac:2022tvc,Luna:2018dpt,Sabharwal:2019ngs,Alawadhi:2020jrv,Godazgar:2020zbv,White:2020sfn,Chacon:2020fmr,Chacon:2021wbr,Chacon:2021hfe,Chacon:2021lox,Dempsey:2022sls,Easson:2022zoh,Chawla:2022ogv,Han:2022mze,Armstrong-Williams:2022apo,Han:2022ubu}
, or
perturbative~\cite{Elor:2020nqe,Farnsworth:2021wvs,Anastasiou:2014qba,LopesCardoso:2018xes,Anastasiou:2018rdx,Luna:2020adi,Borsten:2020xbt,Borsten:2020zgj,Goldberger:2017frp,Goldberger:2017vcg,Goldberger:2017ogt,Goldberger:2019xef,Goldberger:2016iau,Prabhu:2020avf,Luna:2016hge,Luna:2017dtq,Cheung:2016prv,Cheung:2021zvb,Cheung:2022vnd,Cheung:2022mix},
where the latter techniques are of high interest due to potential
applications in gravitational wave physics. Non-perturbative aspects
of the double copy have been explored in
refs.~\cite{Monteiro:2011pc,Borsten:2021hua,Alawadhi:2019urr,Banerjee:2019saj,Huang:2019cja,Berman:2018hwd,Alfonsi:2020lub,Alawadhi:2021uie,White:2016jzc,DeSmet:2017rve,Bahjat-Abbas:2018vgo,Cheung:2022mix,Borsten:2022vtg}. Recent
comprehensive reviews of the double copy can be found in
refs.~\cite{Borsten:2020bgv,Bern:2019prr,Adamo:2022dcm,Bern:2022wqg,White:2021gvv}.

For classical solutions, two incarnations of the double copy have been
particularly well-studied. The first, the Kerr-Schild double copy of
ref.~\cite{Monteiro:2014cda} (see also
refs.~\cite{Didenko:2008va,Didenko:2009td,Didenko:2022qxq}) uses the
traditional tensorial formalism of field theory, and states that
certain exact pure gravity solutions in position space can be written
as simple products of kinematic information entering corresponding
gauge theory solutions. This applies to a special class of solutions
which happen to linearise their respective field theories, where a
particular gauge is picked out (corresponding to the use of
Kerr-Schild coordinates in gravity). A second exact classical double
copy, relevant for the present paper, is the Weyl double copy of
ref.~\cite{Luna:2018dpt}. This uses the spinorial formalism of field
theory, and states that the gauge-independent Weyl spinor in General
Relativity can be written as a certain combination of electromagnetic
spinors, and a scalar field (see
refs.~\cite{Walker:1970un,Hughston:1972qf} for an earlier incarnation
of this idea). This has been shown to work for abitrary vacuum Petrov
type D solutions, as well as some type N solutions relevant for
gravitational waves~\cite{Godazgar:2020zbv}. For solutions of
linearised gauge / gravity theory,
refs.~\cite{White:2020sfn,Chacon:2021wbr} found examples of other
Petrov type, although argued that a generalisation of the Weyl double
copy is needed for Petrov types I and III. Recently, the Weyl double
copy has also been extended to describe non-vacuum
solutions~\cite{Easson:2021asd,Easson:2022zoh}.

It is an ongoing question to establish the general scope and validity
of Weyl double copy-like formulae. We may also ask where they come
from, and a number of works have recently shed light on this
issue. Firstly, refs.~\cite{White:2020sfn,Chacon:2021wbr} formulated
the double copy in twistor space. Certain ``functions'' in the latter
are mapped to spacetime spinor fields by a formula known as the
Penrose transform, and it was shown that a given product of
twistor-space functions exactly reproduces the Weyl double copy in
position space. A conceptual difficulty arises, however, in that the
quantities that enter the Penrose transform can be subjected to
certain equivalence transformations that leave the spacetime fields
invariant. They are thus, strictly speaking, representatives of
cohomology classes rather than functions, and as such cannot usually
be meaningfully multipled together. This is not a problem for deriving
the Weyl double copy in practice, but it does necessitate a
prescription for picking out ``special'' representatives in twistor
space, such that the correct position-space double copy is
obtained. This was addressed further in
refs.~\cite{Adamo:2021dfg,Chacon:2021lox}, both of which gave suitable
prescriptions, albeit with the irksome deficiency that neither choice
obviously corresponds with the original twistor double copy of
refs.~\cite{White:2020sfn,Chacon:2021wbr}. Finally, the situation was
clarified recently in ref.~\cite{Luna:2022dxo}, using ideas developed
in
refs.~\cite{Crawley:2021auj,Guevara:2021yud,Kosower:2018adc,Monteiro:2020plf}. The
latter references show that certain classical gauge / gravity
solutions in position space~\footnote{As is made clear in
refs.~\cite{Guevara:2021yud,Monteiro:2020plf} and below, these
classical solutions are in (2,2) signature, rather than the
conventional Minkowski spacetime with (1,3) signature. The latter can
be obtained via analytic continuation of the coordinates.} can be
obtained as on-shell inverse Fourier transforms of momentum-space
three-point amplitudes. Reference~\cite{Guevara:2021yud} then split
such a transform into two stages, where the first maps amplitudes to
twistor space, and the second corresponds to the Penrose transform
from twistor to position space. The first stage turns out to
correspond to a Laplace transform of the amplitude in energy, and this
procedure will necessarily pick out a certain cohomology
representative in twistor space. As shown in ref.~\cite{Luna:2022dxo},
the representatives obtained for standard three-point amplitudes are
precisely those used in the original twistor double copy of
refs.\cite{White:2020sfn,Chacon:2021wbr}. Thus, the BCJ double copy
for 3-point amplitudes, the twistor double copy, and the
position-space Weyl double copy amount to exactly the same thing,
where overlap exists. Reference~\cite{Luna:2022dxo} also showed that
the simple product-like form of the double copy in twistor space
crucially relies on key properties of three-point amplitudes. In turn,
exact position-space double copies are not expected to be
generic. Thus, the twistor methods provide a way both to
systematically derive Weyl double copy formulae, and to ascertain
their scope.

An interesting analogue of the Weyl double copy was recently
discovered in three spacetime dimensions. Dubbed the {\it Cotton
  double copy}, it relates classical solutions of topologically
massive gauge and gravity theory, in the appropriate spinorial
language~\cite{Emond:2022uaf,Gonzalez:2022otg}. Unlike the 4d Weyl
double copy, the Cotton double copy was found to only work for
solutions of Petrov type N. Reference~\cite{CarrilloGonzalez:2022ggn}
confirmed this by applying similar twistor methods to the
four-dimensional case of ref.~\cite{Luna:2022dxo}, but now in a
three-dimensional context. The appropriate twistor space in this case
is called {\it minitwistor space}, and an appropriate generalisation
of the Penrose transform must be used due to the presence of the
topological mass. By applying similar methods to those developed in
refs.~\cite{Guevara:2021yud,Kosower:2018adc,Monteiro:2020plf},
ref.~\cite{CarrilloGonzalez:2022ggn} showed that this massive Penrose
transform indeed arises naturally upon inverse Fourier transforming
three-point amplitudes, entirely independently of its original
presentation in the twistor literature~\cite{tsai_1996}. The success
of such methods in again deriving and constraining Weyl double-copy
like formulae suggests that similar methods be used to look for
possible Weyl double copies in cases that have not previously been
considered.

With this motivation, we will here consider ${\cal N}=0$ supergravity,
also known as NS-NS gravity. This theory first arose as the effective
field theory emerging in the low energy limit of closed bosonic string
theory (see e.g. refs.~\cite{Green:1987sp,Polchinski:1998rq} for
textbook treatments). As such, it is the theory that is formally
related to pure Yang-Mills theory by the double copy. Indeed, previous
classical double copy formalisms have had to explain why solutions in
pure gravity are obtained by the double copy, when one naturally
expects additional degrees of freedom in the full ${\cal N}=0$ theory
to be turned
on~\cite{Monteiro:2014cda,Goldberger:2016iau,Luna:2017dtq,Luna:2016hge,Kim:2019jwm}. The
latter comprise a scalar -- the {\it dilaton} -- and a two-form field
equivalent to a pseudo-scalar known as the {\it axion} in four
spacetime dimensions. The situation appears to be that for classical
solutions at linear order, one can choose whether or not one sources
the dilaton and / or axion in the gravity theory. However, for
higher-order classical and / or quantum corrections, one must
introduce additional procedures to remove the non-gravitational
degrees of freedom, should they be unwanted. This is indeed the most
common situation for gravitational wave physics, but preserving the
full spectrum offers the chance to ask conceptual questions about the
double copy. In particular, exact position-space classical double
copies for the full ${\cal N}=0$ theory are almost completely
unexplored~\footnote{Very interesting early work in this area set up a
Kerr-Schild ansatz in the framework of double field
theory~\cite{Lee:2018gxc,Cho:2019ype}, although the solutions
considered in this paper are distinct from this
formalism.}. Furthermore, given that we now have techniques for
systematically deriving Weyl double copy formulae, it surely makes
sense to apply these to ${\cal N}=0$ supergravity, and then see what
happens.

A number of recent studies have provided further inspiration. First,
ref.~\cite{Monteiro:2021ztt} used similar methods to
ref.~\cite{Monteiro:2020plf} to write position-space solutions of
${\cal N}=0$ supergravity in terms of on-shell inverse Fourier
transforms of momentum-space scattering amplitudes. They examined
variants of the well-known Kerr and Taub-NUT solutions in pure
gravity, but such that the dilaton and / or axion are also turned
on. A direct consequence of this is that the solutions thus obtained
are no longer vacuum solutions in gravity, and thus a greater variety
of spinor fields in position space are needed, than in the traditional
Weyl double copy of ref.~\cite{Luna:2018dpt}. The authors converted
all spinor fields to the tensorial language, proving that the Riemann
tensor can be written, in general, as a convolution of gauge-theoretic
field strength tensors, and an inverse scalar field. An exact
position-space double copy arises if this convolution is equivalent to
a product, and ref.~\cite{Monteiro:2021ztt} showed explicitly how this
occurs in the tensor language for pure gravity, but also that it is
apparently broken when the dilaton and axion are present. Another
source of inspiration is the recent work of
refs.~\cite{Easson:2021asd,Easson:2022zoh}, which has sought to extend
the Weyl double copy in pure gravity to encompass non-vacuum
solutions. The authors looked at solutions whose source currents /
energy-momentum tensors can be written as a sum of terms which can
each be meaningfully identified across different theories. Then,
Weyl-double-copy-like formulae were proposed for the various spinor
fields that enter the spinorial decomposition of the Riemann
tensor. We will be able to compare our formulae in what follows with
these conjectures.

The aim of this paper is to apply the twistor methods of
refs.~\cite{Guevara:2021yud,Luna:2022dxo,CarrilloGonzalez:2022ggn} to
the starting point of ref.~\cite{Monteiro:2021ztt}. That is, we will
look at the position-space spinor fields generated by the inverse
on-shell Fourier transforms of three-point amplitudes in ${\cal N}=0$
supergravity, and split the transform into two stages, where the first
takes us from momentum to twistor-space. This additional step will
allow us to reveal exact relationships between spinor fields in
position-space that appear to have been overlooked in
ref.~\cite{Monteiro:2021ztt}. For the case of dilaton-axion solutions
with no NUT charge or spin, they indeed take the form of local
products of spinor fields in position space, albeit dressed by
additional factors that mean that they are not of the traditional
Weyl-double copy form. Furthermore, whilst one may write similar
formulae for the case of non-vanishing spin / NUT charge, they do not
have a straightforward double copy interpretation, thus providing
additional insights into the results of
ref.~\cite{Monteiro:2021ztt}. We also compare our results with the
proposed non-vacuum Weyl double copy formulae of
ref.~\cite{Easson:2021asd,Easson:2022zoh}. For the cases we look at,
we do not arrive at the same results. This is not a problem, given
that we are examining a different situation in a different
theory. However, this comparison perhaps suggests that similar methods
to those used in this paper might also prove fruitful in examining the
non-vacuum pure-gravity case.

The structure of our paper is as follows. In section~\ref{sec:review},
we review relevant details of the spinorial formalism, as well as the
arguments of
refs.~\cite{Monteiro:2020plf,Monteiro:2021ztt,Guevara:2021yud,Luna:2022dxo}
which will be needed in what follows. In section~\ref{sec:JNW}, we
apply our twistor methods to the case of a ${\cal N}=0$ supergravity
solution with no spin or NUT charge, finding our first spinorial
double copy formula in position space. In section~\ref{sec:spin}, we
extend the analysis to include the effects of non-zero NUT charge and
spin. Finally, we discuss our results and conclude in
section~\ref{sec:discuss}.

\section{Review of key ideas}
\label{sec:review}

\subsection{The spinorial formalism}
\label{sec:spinor}

Throughout this paper, we will use the spinorial formalism of field
theory, in which all field equations can be written in terms of
two-component Weyl spinors $\pi_A$, and conjugate spinors
$\omega_{\dot{A}}$. Indices may be raised and lowered using the
two-dimensional Levi-Civita symbol:
\begin{equation}
  \pi_A=\epsilon_{BA}\pi^B,\quad \pi^B=-\pi_A \epsilon^{AB},
  \label{piAlower}
\end{equation}
where
\begin{equation}
  \epsilon_{AB}\epsilon^{CB}=\delta^C_A,\quad \epsilon_{01}=1,
  \label{epsilondef}
\end{equation}
and similarly for the dotted equivalent
$\epsilon^{\dot{A}\dot{B}}$. Any tensorial quantity may be converted
into a (multi-index) spinor by contracting spacetime indices its
spacetime indices\footnote{Throughout, we use lower-case Latin
letters, capital Latin letters and Greek letters to correspond to
Lorentz, spinor and twistor indices respectively.} with the {\it
  Infeld-van-der-Waerden symbols}
\begin{equation}
  \sigma^a_{A \dot{A}}=(\mathrm{I},i\sigma_y,\sigma_z,\sigma_x),
  \label{sigmamudef}
\end{equation} 
expressed in terms of the Pauli matrices
\begin{equation}
  \sigma_x=\left(\begin{array}{cc} 0& 1 \\ 1 & 0\end{array}\right),\quad
  \sigma_y=\left(\begin{array}{cc} 0& -i \\ i & 0\end{array}\right),\quad
  \sigma_z=\left(\begin{array}{cc} 1& 0 \\ 0 & -1\end{array}\right).    
  \label{Pauli}
\end{equation}
We have here matched conventions with
refs.~\cite{Monteiro:2020plf,Monteiro:2021ztt}, which work in (2,2)
signature, for reasons that will be clarified below. As an example,
the spinorial translation of a 4-vector is
\begin{equation}
  V_{A\Ad}\equiv V_a \,\sigma^a_{A\Ad}=
  \left(\begin{array}{cc}V_0+V_2 & V_1+V_3\\
    V_3-V_1 & V_0-V_2\end{array}\right).
    \label{Vspinor}
\end{equation} 
The determinant of this matrix is
\begin{equation}
  |V_{A\Ad}|=\left(V_0^2+V_1^2-V_2^2-V_3^2\right)=V^2,
  \label{detV}
\end{equation} 
which vanishes for null vectors, such that for the latter one may
decompose eq.~(\ref{Vspinor}) into an outer product of a spinor and
conjugate spinor:
\begin{equation}
  V_{A\Ad}=\pi_A \tilde{\pi}_{\Ad},\quad V^2=0.
\label{Vfac}
\end{equation}
For later use, we also note the formula
\begin{equation}
  V\cdot W=\frac{1}{2}V_{A\Ad} W^{A\Ad}.
  \label{VdotW}
\end{equation}
The widespread use of the spinorial formalism relies on the fact that
it makes certain structures manifest, that are more difficult to see
in the tensorial framework. This simplification relies on two key
properties, both of which ultimately arise from the fact that each
spinor index may assume one of only two values. The first property is
that all multi-index spinor objects can be decomposed into sums of
products of symmetric spinors, and Levi-Civita symbols. Relevant for
this paper is the spinorial translation of the field strength spinor
$F_{ab}$ in (linearised) gauge theory:
\begin{equation}
  F_{ab}\rightarrow F_{A\dot{A}B\dot{B}}=\phi_{AB}\epsilon_{\dot{A}\dot{B}}+
  \tilde{\phi}_{\dot{A}\dot{B}}\epsilon_{AB},
  \label{Fmunu}
\end{equation}
where $\phi_{AB}$ and $\tilde{\phi}_{\dot{A}\dot{B}}$ are symmetric in
their indices, and represent the self-dual and anti-self-dual parts of
the field respectively. In ${\cal N}=0$ supergravity,
ref.~\cite{Monteiro:2021ztt} introduced a generalised Riemann tensor
$\mathfrak{R}_{abcd}$ for ${\cal N}=0$ supergravity, whose components
represent the combined graviton, dilaton and axion. Its spinorial
translation is
\begin{align}
\mathfrak{R}_{A\Ad B\Bd C\Cd D\Dd}
&=
\mathbf{X}_{ABCD}\,\depsilon_{\Ad\Bd}\,\depsilon_{\Cd\Dd}
+\tilde{\mathbf{X}}_{\Ad\Bd\Cd\Dd}\,\epsilon_{AB}\,\epsilon_{CD} \\
& +\mathbf{\Phi}_{AB\Cd\Dd}\,\depsilon_{\smash{\Ad\Bd}}\,\epsilon_{CD}
+\tilde{\mathbf{\Phi}}_{\Ad\Bd CD}\,\epsilon_{AB}\,\depsilon_{\smash{\Cd\Dd}}~,
\label{eq:curlyRspinor}
\end{align} 
which is directly analogous to the usual spinor decomposition of the
Riemann tensor in General Relativity (see
e.g. refs.~\cite{Penrose:1987uia,Stewart:1990uf}). For vacuum
solutions in pure gravity, the mixed-index spinors are absent, and the
quantity ${\bf X}_{ABCD}$ becomes known as the {\it Weyl spinor},
which we will denote in that context by $\Phi_{ABCD}$. As first
presented in ref.~\cite{Luna:2018dpt}, the Weyl spinors for certain
solutions (those of Petrov type D) can be expressed as a symmetrised
product of gauge theory field strength spinors, divided by a scalar
field $S(x)$:
\begin{equation}
  \Phi_{ABCD}=\frac{\Phi_{(AB}\Phi_{CD)}}{S},\quad
    \tilde{\Phi}_{\Ad\Bd\Cd\Dd}=\frac{\tilde{\Phi}_{(\Ad\Bd}
      \tilde{\Phi}_{\Cd\Dd)}}{\bar{S}},
  \label{WeylDC}
\end{equation}
where $\bar{S}$ is the complex conjugate of $S$ (in Lorentzian
signature), and the two Weyl spinors represent the self-dual and
anti-self-dual graviton degrees of freedom respectively. This is the
{\it Weyl double copy}, and it is the potential generalisation of
these formulae to ${\cal N}=0$ supergravity that we are seeking in
this paper.

\subsection{Twistors}
\label{sec:twistreview}

Twistor theory is a well-established framework combining elements of
algebraic geometry, topology and complex analysis (see
e.g.~\cite{Penrose:1986ca,Huggett:1986fs,Adamo:2017qyl} for reviews),
that has become increasingly prevalent in contemporary research on
scattering amplitudes. One way of introducing twistors is as solutions
of the {\it twistor equation}
\begin{equation}
\nabla_{\dot{A}}^{(A}\Omega^{B)}=0,
\label{twistoreq}
\end{equation} 
where $\nabla_{A\Ad}$ is the spinorial translation of the spacetime
covariant derivative, and $\Omega^B$ a spinor field. The general
solution of this equation in Minkowski space is
\begin{equation}
\Omega^A=\omega^A-x^{A\dot{A}}\pi_{\dot{A}},
\label{twistorsol}
\end{equation}  
such that each solution can be characterised by a four-component
{\it twistor}, containing two spinors:
\begin{equation}
Z^\alpha=\left(\omega^A,\pi_{\dot{A}}\right).
\label{twistor}
\end{equation}
Twistor space $\mathbb{T}$ then consists of the set of all such
objects, and we may define a map from twistor space to spacetime by
defining the ``location'' of a twistor in Minkowksi space to be the
locus of points such that the spinor field $\Omega^A$ vanishes. This
implies the {\it incidence relation}
\begin{equation}
\omega^A=x^{A\dot{A}}\pi_{\dot{A}},
\label{incidence}
\end{equation} 
which is invariant under simultaneous rescalings of both sides:
\begin{equation}
\omega^A\rightarrow \lambda \omega^A,\quad \pi_{\dot{A}}\rightarrow \lambda
\pi_{\dot{A}},\quad \lambda \in\mathbb{C}.
\label{rescale}
\end{equation}
Consequently, twistors obeying the incidence relation constitute
points in {\it projective twistor space} $\mathbb{PT}$. The map
between spacetime and twistor space is non-local: it may be shown that
a point $x^{A\Ad}$ in spacetime maps to a Riemann sphere $X$ in
$\mathbb{PT}$. One way to appreciate this is to note that
eq.~(\ref{incidence}) implies that, upon knowing $x^{A\Ad}$, a twistor
is entirely fixed by the single 2-spinor $\pi_{\Ad}$. Given that the
latter is only defined projectively, we may parametrise all possible
twistors satisfying the incidence relation for a given spacetime point
using the two independent parametrisations
\begin{equation}
  \pi_{\Ad}=(1,\xi),\quad {\rm or}\quad \pi_{\Ad}=(\eta,1),\quad
  \xi,\eta\in\mathbb{C}.
  \label{twistparams}
\end{equation}
The parameters $\xi$ and $\eta$ then indeed correspond to two
conventional coordinate patches for a Riemann sphere, defined via
stereographic projection from the north or south pole to a complex
plane through the equator. On the overlap, one identifies
$\xi\sim\eta^{-1}$.

A key result of twistor theory known as the {\it Penrose transform}
states that solutions of the linearised vacuum field equations for a
spinor field of spin $n$ can be obtained via the contour integral
\begin{equation}
  \tilde{\Phi}_{\Ad_1\Ad_2\ldots\Ad_{2n}}(x)=\frac{1}{2\pi i}\oint_\Gamma
  \pi_{\Ed}d\pi^{\Ed}\,\pi_{\Ad_1}\pi_{\Ad_2}\ldots \pi_{\Ad_{2n}}
  \rho_x[f(Z^\alpha)],
  \label{Penrose}
\end{equation}
where $f(Z^\alpha)$ is a holomorphic function of twistor variables,
$\rho_x$ denotes restriction to the Riemann sphere $X$ corresponding
to the spacetime point $x^a$, and the contour $\Gamma$ on $X$ is such
that it separates any poles of the function $f(Z^\alpha)$. For this
integral to make sense as being in {\it projective} twistor space, the
function $f(Z^\alpha)$ must scale as
\begin{equation}
  f(\lambda Z^\alpha)=\lambda^{-2n-2} f(Z^\alpha).
\label{fscale}
\end{equation}
Furthermore, the quantities $f(Z^\alpha)$ entering the Penrose
transform are not, strictly speaking, functions. One is free to
redefine them according to the equivalence transformations:
\begin{equation}
  f(Z^\alpha)\rightarrow f(Z^\alpha)+f_N(\alpha)+f_S(Z^\alpha),
  \label{fequiv}
\end{equation}
where $f_N(Z^\alpha)$ ($f_S(Z^\alpha)$) has poles only in the northern
(southern) hemisphere of the Riemann sphere $X$. Such contributions
will vanish when performing the Penrose transform, due to having poles
on only one side of the contour $\Gamma$. This infinite freedom to
redefine $f(Z^\alpha)$ is stated more formally by referring to this
quantity as a representative of a cohomology
class~\cite{Eastwood:1981jy}, such that different representatives of
the same cohomology class lead to the same spacetime field.

Note that the Penrose transform of eq.~(\ref{Penrose}) gives only the
anti-self-dual part of the field. One may obtain the self-dual part,
given by an undotted spinor field, in different ways. One way is to
consider the complex conjugate of the twistor equation of
eq.~(\ref{twistoreq})
\begin{equation}
\nabla_A^{(\dot{A}}\Lambda^{\dot{B})}=0,
\label{twistoreq2}
\end{equation}
for which the general solution is:
\begin{equation}
\Lambda^{\dot{A}}=\mu^{\dot{A}}-x^{A\dot{A}}\lambda_{A}.
\label{twistorsol2}
\end{equation}
We may then combine the spinors appearing here to form a {\it dual
  twistor}
\begin{equation}
W_\alpha=\left(\lambda_{A},\mu^{\dot{A}}\right),
\label{dualtwistor}
\end{equation}
satisfying the incidence relation
\begin{equation}
\mu^{\dot{A}}=x^{A\dot{A}}\lambda_A.
\label{incidence2}
\end{equation}
An inner product exists between (dual) twistors:
\begin{equation}
Z^\alpha W_\alpha=\omega^A\lambda_A+\mu^{\dot{A}}\pi_{\dot{A}}.
\label{twistorprod}
\end{equation}
Furthermore, the analogue of the Penrose transform of
eq.~(\ref{Penrose}) in dual twistor space is
\begin{equation}
  \tilde{\Phi}_{A_1A_2\ldots A_{2n}}(x)=\frac{1}{2\pi i}\oint_\Gamma
  \lambda_{E}d\lambda^E\,
  \lambda_{A_1}\lambda_{A_2}\ldots
  \lambda_{A_{2n}}
  \rho_x[f(W_\alpha)],
  \label{Penrose2}
\end{equation}
where again $\Gamma$ is a contour on the Riemann sphere $X$ in
projective dual twistor space corresponding to the spacetime point
$x^a$. To the uninitiated, the above concepts will be highly abstract,
and we refer such a reader to detailed reviews for more
details~\cite{Penrose:1986ca,Huggett:1986fs,Adamo:2017qyl}. The
relevance for the present study is that we will see the above Penrose
transforms emerging naturally upon obtaining classical solutions in
position space from momentum-space scattering amplitudes.

\subsection{Spinor fields from amplitudes}
\label{sec:KMOC}

Reference~\cite{Kosower:2018adc} introduced a systematic method for
obtaining classical solutions from scattering amplitudes, which has
become known as the {\it KMOC formalism}. This was subsequently used
in refs.~\cite{Monteiro:2020plf,Monteiro:2021ztt} to construct
linearised solutions in pure gravity and ${\cal N}=0$
supergravity. More specifically, the specific spinor fields entering
eq.~(\ref{eq:curlyRspinor}) can be obtained from on-shell inverse
Fourier transforms of three-point amplitudes in (2,2) signature, where
the reason for the latter is so that the relevant amplitudes are
non-vanishing once all kinematic constraints are satisfied. Thus
obtained, the fields can be analytically continued to conventional
Lonrentzian (or indeed any other) signature. Let us define three-point
amplitudes for a massive source particle emitting gravitons of a given
helicity ($h^\pm$), dilaton ($\phi$) or axion ($B$) radiation, with
amplitudes ${\cal M}_X$, $X\in\{h,\phi,B\}$. Following
ref.~\cite{Monteiro:2021ztt}, we may define an alternative basis
$\{{\cal M}_{\eta_1\eta_2}\}$ of amplitudes, where
$\eta_i\in\{+1,-1\}$, and such that the physical amplitudes are given
by
\begin{align}
  {\cal M}_{h^{\pm}}&={\cal M}_{\pm\pm};\\
  {\cal M}_{\phi}&=\frac12\Big(
  {\cal M}_{+-} + {\cal M}_{-+}\Big);\notag\\
  {\cal M}_{B}&=\frac12\Big(
  {\cal M}_{+-} - {\cal M}_{-+}\Big).
  \label{M+-def}
\end{align}
The motivation for this definition is the double copy for scattering
amplitudes, which states that given degrees of freedom in ${\cal N}=0$
supergravity can be obtained as certain combinations of polarisation
states from two separate gauge theories. In its simplest form, this
holds for plane wave states, where it is known that the polarisation
tensors for gravitons of given helicity can be obtained as outer
products of photon polarisation vectors:
\begin{equation}
  \epsilon_{ab}^\pm=\epsilon_a^\pm\epsilon_b^\pm.
  \label{EM1}
\end{equation}
Likewise, the two independent combinations
\begin{equation}
  \frac12\left(\epsilon_a^+\epsilon_b^- \pm
  \epsilon_a^-\epsilon_b^+ \right)
  \label{EM2}
\end{equation}
turn out to describe the dilaton and axion respectively. The
non-trivial statement of the double copy is that this product
structure survives even when interactions are included, so that
multi-point amplitudes involving gravitons, axions and dilatons can be
obtained by combining gauge theory amplitudes in appropriate
combinations. For three-point amplitudes involving radiation of a
single $(h^\pm,\phi,B)$ state this is very simple, and corresponds
precisely to eq.~(\ref{M+-def}), where the amplitudes ${\cal
  M}_{\eta_1\eta_2}$ are given by simple products of gauge theory
three-point amplitudes with the appropriate helicity. Motivated by
this, ref.~\cite{Monteiro:2021ztt} proposes a slightly more general
relationship
\begin{align} 
\label{eq:DCmap}
\ampM_{\hel_L \hel_R}
=
-\frac{\kappa}{4\,Q^2}\,c_{\hel_L \hel_R}\,\ampA^{(L)}_{\hel_L}\,\ampA^{(R)}_{\hel_R}~,
\end{align}
where $\kappa=\sqrt{32\pi G_N}$ is the gravitational coupling constant
in terms of the Newton constant $G_N$, and $Q$ the electromagnetic
coupling of the source particle in the gauge theory. We also adopt the
notation of ref.~\cite{Monteiro:2021ztt}, such that the two gauge
theories that are double-copied to make the gravity theory are
referred to as the ``left'' ($L$) and ``right'' ($R$) theories.  The
additional constants $\{c_{\eta_L\eta_R}\}$ allow different
normalisations for different helicity combinations, where the
traditional BCJ double copy has these all equal to one.

Using the so-called {\it KMOC formalism} of
ref.~\cite{Kosower:2018adc}, ref.~\cite{Monteiro:2021ztt} showed that
the classically observed values for the spinor fields appearing in
eq.~(\ref{eq:curlyRspinor}) at linearised level -- derived as
expectation values of quantum field operators -- are given in terms
of the above amplitudes via
\begin{align}
{\mathbf{X}}_{ABCD}
&=
-\frac{\kappa^2c_{++}}{2Q^2}\,\Re\,i\int\d \Phi(k)  \del(2 p \cdot k) \,
	\ampA_+^{(L)}\ampA_+^{(R)}\,
\ket{k}_A\ket{k}_B
\ket{k}_C\ket{k}_D\,e^{-ik\cdot x}~ \label{posgraviton},
\\[1em]
\tilde{\mathbf{X}}_{\Ad\Bd\Cd\Dd}
&=
-\frac{\kappa^2c_{--}}{2Q^2}\,\Re\,i\,\int\d \Phi(k) \del(2 p \cdot k) 
 \,
	\ampA_-^{(L)}\ampA_-^{(R)}\,
[k|_{\Ad} [k|_{\Bd}
[k|_{\Cd}[k|_{\Dd}\,e^{-ik\cdot x}~ \label{neggraviton},
\\[1em]
{\mathbf{\Phi}}_{AB\Cd\Dd}
&=
+\frac{\kappa^2c_{+-}}{2\,Q^2}\,\Re i\int\d\Phi(k) \del(2 p \cdot k) 
 \,
	\ampA_+^{(L)}\ampA_-^{(R)}\,
\,\ket{k}_A\ket{k}_B[k|_{\Cd}[k|_{\Dd}\,e^{-ik\cdot x} \label{da}~,
\\[1em]
{\tilde{ \mathbf{\Phi}}}_{\Ad\Bd CD}
&=
+\frac{\kappa^2c_{-+}}{2\,Q^2}\,\Re i\int\d\Phi(k) \del(2 p \cdot k)  \,
	\ampA_-^{(L)}\ampA_+^{(R)}\,
\,[k|_{\Ad}[k|_{\Bd} \ket{k}_C\ket{k}_D\,e^{-ik\cdot x}~\label{conda}. 
\end{align} 
Here we have introduced the Lorentz-invariant phase space measure
\begin{equation}
  d\Phi(k)=\frac{d^4 k}{(2\pi)^4}\hat{\delta}(k^2)\Theta(k^0),
  \label{dPhik}
\end{equation}
the normalised delta function
\begin{align}
\hat{\delta}(x) = 2 \pi \delta(x), \label{delta}
\end{align}
and the conventional bra-ket notation whereby the spinorial
translation of the null radiation momentum $k^a$ is written as
\begin{equation}
  k_{A\Ad}=\ket{k}_A [k|_{\Ad}.
  \label{kbraket}
\end{equation}
In words, eqs.~(\ref{posgraviton}--\ref{conda}) constitute the fact
that classical spinor fields in ${\cal N}=0$ supergravity at
linearised level can be expressed in terms of on-shell inverse Fourier
transforms of three-point amplitudes, where the latter are obtained by
double-copying gauge theory amplitudes according to
eq.~(\ref{eq:DCmap}). Reference~\cite{Monteiro:2021ztt} used these
formulae as a starting point for examining whether or not one may
formulate exact position-space double copies for particular solutions,
after converting all spinor fields into the tensorial language. We
will consider the same family of solutions here, namely those
constructed from gauge theory amplitudes
\begin{align}
\ampA^{(L,R)}_\eta=-2Q(p\cdot\epsilon_\eta)e^{\hel(\theta_{L,R}+ik\cdot
  a_{L,R})}, \label{photonamp}
\end{align} 
for some vectors $\{a_{R,L}\}$ and constant parameters $\theta_{L,R}$. Here
\begin{equation}
  p^a=M u^a
  \label{pmu}
\end{equation}
is the 4-momentum of the source, with mass $M$ and 4-velocity
$u^a$. We have also introduced polarisation vectors for the photon
in each theory, explicit realisations of which are given by (see
e.g. ref.~\cite{Monteiro:2020plf})
\begin{equation}
  \epsilon_{-}^a=(\epsilon_+^a)^\ast=
  -\frac{\langle k|\sigma^a|l]}{\sqrt{2}[kl]},
  \label{poldef}
\end{equation}
where $|l]^{\Ad}$ is an arbitrary null reference spinor, corresponding
  to a gauge choice. The physical interpretation of the amplitudes in
  eq.~(\ref{photonamp}) has been explored in detail in
  ref.~\cite{Emond:2020lwi}. Without the exponential factor (i.e. for
  $a^a=\theta=0$ in a given gauge theory), the amplitude describes
  photon emission from a spinless particle with electric charge, and
  thus gives rise to the Coulomb solution. The double copy of this in
  pure gravity is the Schwarzschild solution~\cite{Monteiro:2014cda},
  and in ${\cal N}=0$ supergravity leads to the {\it JNW
    solution}~\cite{Janis:1968zz}, in which a static,
  spherically-symmetric black hole is dressed by a non-zero dilaton
  profile. Upon turning on the vector $a^a$ in gauge theory or pure
  gravity, a spin is generated for the source particle, such that
  $a^a$ can be identified with its Pauli-Lubanski spin
  pseudo-vector. In pure gravity the resulting field is that of the
  {\it Kerr solution}~\cite{Kerr:1963ud}, and its gauge theory single
  copy is known as $\sqrt{\rm Kerr}$. The exponential factor appearing
  in eqs.~(\ref{photonamp}) then constitutes the {\it Newman-Janis
    shift}, a complex transformation that transforms the Schwarzschild
  into the Kerr black hole~\cite{Newman:1965tw}. The additional
  constant factor in the exponent transforms to a different solution
  in pure gravity, namely the {\it NUT solution} of
  refs.~\cite{Taub,NUT}. This is a generalisation of the Schwarzschild
  black hole that has a non-vanishing rotational character to the
  gravitational field at asymptotic infinity. Its single copy is a
  dyon, where the NUT charge in the gravity theory maps to the
  magnetic monopole charge in the gauge theory~\cite{Luna:2015paa}. In
  ${\cal N}=0$ supergravity, double-copying the amplitudes of
  eq.~(\ref{photonamp}) will thus lead to generalisations of the JNW
  black hole, in which both spin and NUT charge are present. This in
  turn entails the possibility that the axion may turn on in addition
  to the dilaton, as discussed in ref.~\cite{Monteiro:2021ztt}.

In the remainder of this paper, we will adopt the starting point of
eqs.~(\ref{posgraviton}--\ref{conda}) as in
ref.~\cite{Monteiro:2021ztt}, but take a different approach in looking
for position-space double copies. That is, we will stick with the
spinorial language, and also use the twistor methods of
refs.~\cite{Guevara:2021yud,Luna:2022dxo,CarrilloGonzalez:2022ggn},
that have recently proved successful in deriving Weyl-double copy
formulae in pure gravity. As a warm-up, we consider the simplest
solution -- corresponding to double-copying the Coulomb charge -- in
the following section.

\section{A spinorial double copy for the JNW solution}
\label{sec:JNW}

\subsection{From momentum to twistor space}
\label{sec:twistor}

In refs.~\cite{Luna:2022dxo,CarrilloGonzalez:2022ggn} (motivated by
ref.~\cite{Guevara:2021yud}), the inverse Fourier transform appearing
in eqs.~(\ref{posgraviton}--\ref{conda}) is split into two stages,
such that one considers an intermediate twistor space between momentum
and position space. This in turn provides additional insights allowing
one to systematically derive position-space double copy formulae, and
to ascertain when they apply. To illustrate this idea, let us focus on
the gauge theory analogue of eqs.~(\ref{posgraviton}--\ref{conda}),
namely the fact that the field strength spinors corresponding to the
amplitudes of eq.~(\ref{photonamp}) are given by
\begin{align}
  \langle{\Phi}_{AB}\rangle
&=
\Re\, \frac{\sqrt{2}}{M}\int\d \Phi(k)  \del(u \cdot k) \,
	\ampA_+ \,
\ket{k}_A\ket{k}_B
\,e^{-ik\cdot x}\,, \label{posphoton}\\
  \langle{\tilde{\Phi}}_{\Ad\Bd}\rangle
&=
\Re\,\frac{\sqrt{2}}{M}\int\d \Phi(k)  \del(u\cdot k) \,
	\ampA_- \,
[k_{\Ad}|\,[k_{\Bd}|
\,e^{-ik\cdot x}~ \label{negphoton}.
\end{align}
In order to carry out the phase-space integral, one may make the
change of variables
\begin{equation}
  k_{A \dot{A}}=\omega \lambda_A\tilde{\lambda}_{\dot{A}}+\xi q_{A\dot{A}},
  \label{kAA'}
\end{equation}
where $q_{A\Ad}$ is the spinorial translation of an arbitrary constant
null 4-vector, and the spinors $\lambda_A$ and $\tilde{\lambda}_{\Ad}$
are defined only up to an overall scaling, which may be absorbed into
$\omega$. The various bra-ket symbols above are then given in these
new variables by
\begin{equation}
  |k\rangle_A=\omega^{1/2}\lambda_A,\quad
  \langle k|^A=\omega^{1/2}\lambda^A,\quad
  [k|_{\Ad}=\omega^{1/2}\tilde{\lambda}_{\Ad},\quad
    |k]^{\Ad}=\omega^{1/2}\tilde{\lambda}^{\Ad}.
  \label{kspinors}
\end{equation}
If we choose to parametrise the spinors via
\begin{equation}
  \lambda_A=(1,z),\quad \tilde{\lambda}_{\Ad}=(1,\tilde{z}),\quad
  z,\tilde{z}\in\mathbb{C},
  \label{lambdaparam}
\end{equation}
then the change of variables is from the four components of $k^a$ to
the set $(\omega,\xi,z,\tilde{z})$. After evaluating the Jacobian, one finds
\begin{align}
d\Phi(k) = \frac{d z d\tilde{z} d\omega d\xi \delta(\xi) \omega}{4 (2
  \pi)^{3}},
\label{measure}
\end{align}
such that the field strength spinor with dotted indices becomes
\begin{equation}
  \tilde{\Phi}_{\Ad\Bd}=\frac{2\sqrt{2}}{4M(2\pi)^2}\Re \int dz
  d\tilde{z}d\omega d\xi\,\delta(\xi) \delta(u\cdot
  k)\omega^2\Theta(\omega){\cal
    A}_-\,\tilde{\lambda}_{\Ad}\tilde{\lambda}_{\Bd} e^{-ik\cdot x}.
\label{PhiABcalc1}
\end{equation}
The $\xi$ integral can be carried out immediately, and simply sets
$\xi=0$. This corresponds to the fact that $\xi$ in eq.~(\ref{kAA'})
parametrises how far $k^a$ is from being null, and thus
on-shell. However, eq.~(\ref{posphoton}) manifestly contains an
on-shell Fourier transform, and thus the vanishing of $\xi$ enforces
this on-shell condition. To go further in carrying out the integral,
let us substitute the explicit form of the Coulomb amplitude, namely
that obtained from eq.~(\ref{photonamp}) by setting
$\theta=a^a=0$. From eq.~(\ref{poldef}) we find
\begin{equation}
  {\cal A}_-= \frac{2MQ}{\sqrt{2}}\frac{u_{A\Ad}\lambda^A l^{\Ad}}
  {\tilde{\lambda}_{\Bd}l^{\Bd}},
  \label{A-form}
\end{equation}
such that eq.~(\ref{PhiABcalc1}) becomes
\begin{align}
  \tilde{\Phi}_{\Ad\Bd}&=\frac{Q}{(2 \pi)^{2}}
  \Re \int dz d\tilde{z}d\omega 
  \delta\left(u_{A\Ad}\lambda^A\tilde{\lambda}^{\Ad}\right)
  \omega\Theta(\omega)
        \tilde{\lambda}_{\Ad}\tilde{\lambda}_{\Bd}
        e^{-\frac{i\omega}{2}\lambda_A\tilde{\lambda}_{\Ad}x^{A\Ad}}\notag\\
        &\quad\times \frac{u_{A\Ad}\lambda^A l^{\Ad}}{\tilde{\lambda}_{\Bd}
        l^{\Bd}}.
\label{PhiABcalc2}
\end{align}
The remaining delta function sets
\begin{equation}
  \lambda_A\propto u_{A\Ad}\tilde{\lambda}^{\Ad} 
\label{lambdaprop}
\end{equation}
which, as emphasised in ref.~\cite{Luna:2022dxo}, can be turned into
an equality by reparametrising
\begin{equation}
  \lambda_A=\left(\frac{1}{\sqrt{z}},\sqrt{z}\right),\quad
  \tilde{\lambda}_{\Ad}=\left(
\frac{1}{\sqrt{-\tilde{z}}},-\sqrt{-\tilde{z}}
\right).
\label{lambdareparam}
\end{equation}
One may then use the delta function to eliminate the $z$ integral,
yielding
\begin{equation}
  \tilde{\Phi}_{\Ad\Bd}=\frac{Q}{(2 \pi)^{2}}
  \Re \int d\tilde{z}\,
        \tilde{\lambda}_{\Ad}\tilde{\lambda}_{\Bd}
        \mathfrak{M}(\tilde{\lambda}_{\Ad}),
        \label{PhiABcalc3}
\end{equation}
where
\begin{align}
  \mathfrak{M}(\tilde{\lambda}_{\Ad})&=\int d\omega\,\omega\, \Theta(\omega)
  e^{-\frac{i\omega}{2}{u_A}^{\Bd}x^{A\Ad}\tilde{\lambda}_{\Ad}
    \tilde{\lambda}_{\Bd}}\notag\\
  &=-\frac{4}{[{u_A}^{\Bd}x^{A\Ad}\tilde{\lambda}_{\Ad}
    \tilde{\lambda}_{\Bd}]^2}.
  \label{Mgothic}
\end{align}
At this stage, we may reparametrise back to the original definition of
$\tilde{z}$. From eq.~(\ref{lambdaparam}), it then follows that
\begin{equation}
  d\tilde{z}=\tilde{\lambda}_{\Ed}d\lambda^{\Ed}
  \label{dtildez}
\end{equation}
is the projective measure on the Riemann sphere parametrised (in a
particular coordinate patch) by $\tilde{z}$. Next, we may note that
$\mathfrak{M}(\tilde{\lambda}_{\Ad})$ depends on
$\tilde{\lambda}_{\Ad}$ through the specific combinations
$\tilde{\lambda}_{\Ad}$ and
\begin{displaymath}
  \omega^A=x^{A\Ad}\tilde{\lambda}_{\Ad},
\end{displaymath}
where we can recognise the incidence relation of
eq.~(\ref{incidence}). Thus, the quantity
\begin{equation}
  Z^\alpha=(\tilde{\lambda}_{\Ad},\omega^A)
  \label{Zalphadef}
\end{equation}
is a bona fide point in projective twistor space $\mathbb{PT}$! We may
also then write the functional dependence in eq.~(\ref{Mgothic}) as
$\mathfrak{M}(\tilde{\lambda}_{\Ad})\equiv
\rho_x[\mathfrak{M}(Z^\alpha)]$, such that eq.~(\ref{PhiABcalc3})
becomes
\begin{equation}
  \tilde{\Phi}_{\Ad\Bd}=-\frac{Q}{(2\pi)^2}\Re 
  \oint \tilde{\lambda}_{\Ed}
  d\tilde{\lambda}^{\Ed} \,\tilde{\lambda}_{\Ad}\tilde{\lambda}_{\Bd}
  \,\rho_x[\mathfrak{M}(Z^\alpha)].
  \label{PhiABcalc4}
\end{equation}
It is now straightforward to recognise the Penrose transform of
eq.~(\ref{Penrose}). Furthermore, the function of eq.~(\ref{Mgothic})
is homogeneous of degree $-4$ under rescalings of $Z^\alpha$ (and thus
$\tilde{\lambda}_{\Ad}$), in agreement with eq.~(\ref{fscale}).

Let us summarise what has happened. We started by expressing spacetime
(spinor) fields as inverse on-shell Fourier transforms of
momentum-space scattering amplitudes. Next, we transformed to spinor
variables, and found out that carrying out ``half'' of the Fourier
transform takes our amplitude into twistor space. Indeed,
eq.~(\ref{Mgothic}) is a variant of the so-called half Fourier
transform used in the seminal work of ref.~\cite{Witten:2003nn}, which
originated the modern use of twistor methods in scattering amplitude
research. Here it takes the form of a Laplace transform in $\omega$,
which from eq.~(\ref{kAA'}) can be interpreted as the energy of the
radiation. Note that, in defining a precise form for
$\mathfrak{M}(Z^\alpha)$, the half transform defines a particular
cohomology representative in twistor space for a given classical
spacetime field i.e. one that is ``picked out'' by the
amplitude. Reference~\cite{Luna:2022dxo} used this observation to
resolve cohomological ambiguities in the twistor double copy of
refs.~\cite{White:2020sfn,Chacon:2021wbr}. 

\subsection{Consistency relation between (anti-)self dual field strength spinors}
\label{sec:relation}

In eq.~(\ref{PhiABcalc4}), we have shown that the anti-self-dual field
strength spinor can be obtained as an explicit Penrose transform of a
cohomology representative derived from a momentum-space scattering
amplitude. One may also perform a similar exercise for the self-dual
spinor, starting from eq.~(\ref{posphoton}), and such that the
analogue of eq.~(\ref{PhiABcalc2}) is found to be
\begin{align}
  \Phi_{AB}&=\frac{Q}{(2 \pi)^{2}}\Re \int dz d\tilde{z}d\omega\delta\left(
  u_{A\Ad}\lambda^A\tilde{\lambda}^{\Ad}\right)\omega\Theta(\omega)
  \lambda_A\lambda_Be^{-\frac{i\omega}{2}\lambda_A\tilde{\lambda}_{\Ad}x^{A\Ad}}
  \notag\\
  &\times \frac{u_{A\Ad}l^A\tilde{\lambda}^{\Ad}}{l^A\lambda_A}.
  \label{PhiABcalcb}
\end{align}
In this case, we can use the delta function to set
\begin{equation}
  \tilde{\lambda}_{\Ad}=u_{A\Ad}\lambda^A,
  \label{lambdaprop2}
\end{equation}
namely the inverse relation of eq.~(\ref{lambdaprop}). This eliminates
the $\tilde{z}$ integral in eq.~(\ref{PhiABcalcb}), such that one gets
\begin{equation}
  \Phi_{AB}=\frac{Q}{(2\pi)^{2}}\Re \int dz \lambda_A\lambda_B
  \mathfrak{N}(\lambda_A),
  \label{PhiABcalcb2}
\end{equation}
where
\begin{align}
  \mathfrak{N}(\lambda_A)&=\int d\omega \omega \Theta(\omega)
  e^{\frac{i\omega}{2}{u^B}_{\Ad}x^{A\Ad}\lambda_A\lambda_B}\notag\\
  &=\frac{-4}{[{u^B}_{\Ad}x^{A\Ad}\lambda_A\lambda_B]^2}.
  \label{Ngothic}
\end{align}
Recognising that this depends only on $\lambda_A$ and
\begin{displaymath}
  \mu^{\Ad}=x^{A\Ad}\lambda_A,
\end{displaymath}
we see that
$\mathfrak{N}(\lambda_A)\equiv\rho_x[\mathfrak{N}(W_\alpha)]$ is
defined on dual twistor space, where $W_\alpha$ is given as in
eq.~(\ref{dualtwistor}). Equation~(\ref{PhiABcalcb2}) then becomes
\begin{equation}
  \Phi_{AB}=\frac{Q}{(2\pi)^{2}}\Re \oint \lambda_Ed\lambda^E\,\lambda_A
  \lambda_B\rho_x[\mathfrak{N}(W_\alpha)],
  \label{PhiABres}
\end{equation}
and we recover the well-known result that anti-self-dual (self-dual)
solutions are associated with Penrose transforms from twistor space
(dual twistor space) respectively. However, a fact that was
overlooked in the recent ref.~\cite{Luna:2022dxo} is that it is also
possible to derive a consistency relation between the (anti-)self-dual
spinor fields, for this particular solution. Returning to
eq.~(\ref{PhiABcalc2}), we may choose to eliminate
$\tilde{\lambda}_{\Ad}$ using the delta function, rather than
$\lambda_A$ i.e. by using eq.~(\ref{lambdaprop2}) rather than
eq.~(\ref{lambdaprop}). We then find
\begin{equation}
  \tilde{\Phi}_{\Ad\Bd}={u^A}_{\Ad}{u^B}_{\Bd}\left[
    \frac{Q}{(2\pi)^{2}}\Re \int dz\,\lambda_A\lambda_B
    \rho_x[\mathfrak{N}(\lambda_A)]\right],
  \label{Phiucontract}
\end{equation}
where $\mathfrak{N}(\lambda_A)$ is defined as in
eq.~(\ref{Ngothic}). In other words, we have obtained the relationship
\begin{equation}
  \tilde{\Phi}_{\Ad\Bd}(x)={u^A}_{\Ad}{u^B}_{\Bd}\Phi_{AB}(x),
  \label{Phirel}
\end{equation}
whose inverse -- as may be derived by using eq.~(\ref{lambdaprop}) in
eq.~(\ref{PhiABcalcb}) -- is
\begin{equation}
  \Phi_{AB}={u_A}^{\Ad}{u_B}^{\Bd}\tilde{\Phi}_{\Ad\Bd}.
  \label{Phirel2}
\end{equation}
It is instructive to see how this relation actually works in practice,
by finding the explicit forms of the spinor fields implied by the
Penrose transforms of eqs.~(\ref{PhiABcalc4}, \ref{PhiABres}). Let us
first note that we may write the combination appearing in
eq.~(\ref{Mgothic}), using the parametrisation of
eq.~(\ref{lambdaparam}), as
\begin{equation}
  {u_A}^{\Bd}x^{A\Ad}\tilde{\lambda}_{\Ad}\tilde{\lambda}_{\Bd}
  =\tilde{N}^{-1}(x)(\tilde{z}-\tilde{z}_1)(\tilde{z}-\tilde{z}_2),
  \label{Nxidef}
\end{equation}
where an explicit calculation yields
\begin{equation}
  \tilde{N}^{-1}(x)=x^{0\dot{1}},\quad \tilde{z}_{1,2}=\frac{\left( x^{1 \dot{1}}- x^{0 \dot{0}}\right) \pm
    \sqrt{(x^{0 \dot{0}})^{2} +(x^{1 \dot{1}})^{2} -2 x^{0 \dot{0}}
      x^{1 \dot{1}}+4x^{0 \dot{1}}x^{1 \dot{0}}}}{2 x^{0 \dot{1}}}.
  \label{tildez12def}
\end{equation}
From eqs.~(\ref{PhiABcalc3}, \ref{Mgothic}), we then have
\begin{equation}
  \tilde{\Phi}_{\Ad\Bd}=\frac{4Q}{4\pi^2}\tilde{N}^2(x)\Re
  \oint d\tilde{z} \frac{(1,\tilde{z})_{\Ad}(1,\tilde{z})_{\Bd}}
        {(\tilde{z}-\tilde{z}_1)^2(\tilde{z}-\tilde{z}_2)^2},
        \label{Pencalc1}
\end{equation}
and we may carry out the $\tilde{z}$ integral by enclosing the pole at
$\tilde{z}=\tilde{z}_1$ (see e.g. ref.~\cite{Chacon:2021wbr} for a similar
calculation). The result is
\begin{equation}
  \tilde{\Phi}_{\Ad\Bd}=-\frac{4Q}{2\pi}\frac{1}{[x^{0\dot{1}}]^2}
  \Re i\left[\frac{\tilde{\alpha}_{(\Ad}\tilde{\beta}_{\Bd)}}
    {(\tilde{z}_1-\tilde{z}_2)^3}
    \right],
  \label{Phitilderesult}
\end{equation}
where the principal spinors are given by
\begin{equation}
  \tilde{\alpha}_{\Ad}=(1,\tilde{z}_1),\quad
  \tilde{\beta}_{\Bd}=(1,\tilde{z}_2).
  \label{principals1}
\end{equation}
Substituting the result of eq.~(\ref{tildez12def}) and simplifying,
one finds
\begin{equation}
  \tilde{\Phi}_{\Ad\Bd}=\frac{4Q}{2\pi}
  \Re i \left[\frac{1}{\Lambda^{3/2}(x)}
  \left(\begin{array}{cc}
    x^{0\dot{1}} & x^{1\dot{1}}-x^{0\dot{0}}\\
    x^{1\dot{1}}-x^{0\dot{0}} & -x^{1\dot{0}}
  \end{array}
  \right)
  \right],
  \label{PhiABres2}
\end{equation}
where
\begin{equation}
    \Lambda=(x^{0\dot{0}})^2+(x^{1\dot{1}})^2-2x^{0\dot{0}}x^{1\dot{1}}
  +4x^{0\dot{1}}x^{1\dot{0}}.
\label{Lambdadef}
\end{equation}
Likewise, one may carry out the Penrose transform in
eq.~(\ref{PhiABres}), and the result is
\begin{equation}
  \Phi_{AB}=-\frac{4Q}{2\pi}
  \Re i \left[\frac{1}{\Lambda^{3/2}(x)}
  \left(\begin{array}{cc}
    x^{1\dot{0}} & x^{1\dot{1}}-x^{0\dot{0}}\\
    x^{1\dot{1}}-x^{0\dot{0}} & -x^{0\dot{1}}
  \end{array}
  \right)
  \right].
  \label{PhiABres3}
\end{equation}
To confirm the relation of eq.~(\ref{Phirel}), we may note that the
static 4-velocity $u^a=(1,\vec{0})$ implies
\begin{equation}
  {u^A}_{\Ad}=\epsilon^{BA}u_{A\Ad}=\left(\begin{array}{cc}
    0 & 1 \\ -1 & 0
  \end{array}\right).
  \label{umat}
\end{equation}
Then we have
\begin{align*}
  {u^A}_{\Ad}{u^B}_{\Bd} \Phi_{AB}=
  -\frac{4Q}{2\pi}
  \Re i \left[\frac{1}{\Lambda^{3/2}(x)}
\left(\begin{array}{cc}
    0 & -1 \\ 1 &0 
  \end{array}\right)
    \left(\begin{array}{cc}
    x^{1\dot{0}} & x^{1\dot{1}}-x^{0\dot{0}}\\
    x^{1\dot{1}}-x^{0\dot{0}} & -x^{0\dot{1}}
  \end{array}
    \right)
\left(\begin{array}{cc}
    0 & 1 \\ -1 & 0
  \end{array}\right)    
  \right],
\end{align*}
which indeed agrees with eq.~(\ref{PhiABres2}). Note that we have here
chosen to express the field-strength spinor in terms of the components
$x^{A\Ad}$, rather than substitute explicit spacetime coordinates. It
is then straightforward to evaluate these formula in either (2,2) or
(1,3) signature.

\subsection{A double copy formula for the JNW solution}
\label{sec:doubleJNW}

We now have all the ingredients we need to ascertain the existence --
or otherwise -- of a double copy formula for the JNW solution. Let us
start with the spinor $\tilde{\bf X}_{\Ad\Bd\Cd\Dd}$, given in terms
of an amplitude by eq.~(\ref{posgraviton}). Carrying out similar steps
to that leading to eq.~(\ref{PhiABcalc4}), we find
\begin{align}
  \tilde{\mathbf{X}}_{\Ad\Bd\Cd\Dd}=  \frac{\kappa^2c_{--} M}
        {(2\pi)^2 } \Re \oint
  \tilde{\lambda}_{\Ed} d\tilde{\lambda}^{\Ed}\,
  \tilde{\lambda}_{\dot{A}} \tilde{\lambda}_{\dot{B}}
\tilde{\lambda}_{\dot{C}} \tilde{\lambda}_{\dot{D}}\,\rho_x
  \left[\frac{4}{(\tilde{\lambda}_{\dot{A}} U\indices{_A ^{\dot{B}}} \tilde{\lambda}_{\dot{B}} x^{A\dot{A}})^{3}}\right],
\label{neggraviton6}
\end{align}  
where we have substituted the explicit form of the amplitudes of
eq.~(\ref{photonamp}), and again used the delta function
$\delta(u\cdot k)$ to impose the condition of
eq.~(\ref{lambdaprop}). Carrying out the Penrose transform using the
parametrisation of eq.~(\ref{lambdaparam}) (see
e.g. ref.~\cite{Chacon:2021wbr} for a similar calculation), one finds
\begin{equation}
  \tilde{\bf X}_{\Ad\Bd\Cd\Dd}=\frac{\kappa^2 c_{--}i M}{2 \pi }\Re\left[
    \frac{\tilde{N}^3(x)}{(\tilde{z}_1-\tilde{z}_2)^5}\tilde{\alpha}_{(\Ad}
    \tilde{\beta}_{\Bd}\tilde{\alpha}_{\Cd}\tilde{\beta}_{\Dd)},
    \right]
  \label{tildeXres}
\end{equation}
where $\tilde{N}(x)$, $\tilde{z}_{1,2}$ and the principal spinors
$\tilde{\alpha}_{\Ad}$ and $\tilde{\beta}_{\Ad}$ have been defined in
eqs.~(\ref{tildez12def}, \ref{principals1}). One may also define a
scalar field
\begin{equation}
  \tilde{S}\propto \frac{N(x)}{\tilde{z}_1-\tilde{z}_2},
  \label{tildeSJNW}
\end{equation}
which can be found to satisfy the massless Klein-Gordon equation. Then
comparison of eq.~(\ref{tildeXres}) with eqs.~(\ref{Phitilderesult},
\ref{tildeSJNW}) implies the relationship
\begin{equation}
  \tilde{\bf X}_{\Ad\Bd\Cd\Dd}=\frac{\tilde{\Phi}_{(\Ad\Bd}
    \tilde{\Phi}_{\Cd\Dd)}}{\tilde{S}},
  \label{DC1}
\end{equation}
where all constant factors have been absorbed into the scalar function
$\tilde{S}(x)$. This is a precise analogue of the Weyl double copy
formula of eq.~(\ref{WeylDC}), and indeed its derivation using twistor
methods is exactly the same as in ref.~\cite{Luna:2022dxo}. Similar
arguments may be used to verify the corresponding relationship
\begin{equation}
  {\bf X}_{ABCD}=\frac{\Phi_{(AB}\Phi_{CD)}}{S},
  \label{DC2}
\end{equation}
where the scalar field $S$ is defined by
\begin{equation}
  S(x)\propto\frac{N(x)}{z_1-z_2},\quad N^{-1}(x)=x^{1\dot{0}}.
  \label{SJNW}
\end{equation}
Given the close analogues of these formulae with their pure gravity
counterparts, it is perhaps not surprising that they occur. A more
interesting question is whether or not a double copy formula emerges
for the mixed-index spinors appearing in eq.~(\ref{eq:curlyRspinor}),
which have no counterpart for vacuum solutions in pure gravity. Let us
first consider
\begin{align}
\langle{\mathbf{\Phi}}_{AB\Cd\Dd}\rangle &= \frac{\kappa^2 c_{+-} M^{2}}{4
  (2 \pi)^{2}}\Re\,i\int dz d\tilde{z} d\omega \omega^{3} \Theta(\omega)
\delta(2 p\cdot k) \,\lambda_{A} \lambda_B
\tilde{\lambda}_{\Cd}\tilde{\lambda}_{\Dd}\,e^{-\frac{i\omega}{2}\lambda_A\tilde{\lambda}_{\dot{A}}x^{A\dot{A}}} \notag\\
&\times \left(\frac{u_{A\Ad}\lambda^Al^{\Ad}}{\tilde{\lambda}_{\Bd}l^{\Bd}}
\right)
\left(\frac{u_{D\Dd}l^D\tilde{\lambda}^{\Dd}}{l^C\lambda_C}\right)
\label{da2}~,
\end{align}
where we have substituted the amplitudes of eq.~(\ref{photonamp}) into
eq.~(\ref{da}). We may then use the delta function to implement either
of the conditions of eq.~(\ref{lambdaprop}) or
eq.~(\ref{lambdaprop2}). Choosing the former, we get
\begin{align}
\langle{\mathbf{\Phi}}_{AB\Cd\Dd}\rangle &= \frac{\kappa^2 c_{+-} M}{
4  (2 \pi)^{2}}{u^C}_{\Cd}{u^D}_{\Dd}\Re\,i 
  \int dz  d\omega \omega^{2} \Theta(\omega)
 \,\lambda_{A} \lambda_B
\lambda_{C}\lambda_{D}\,e^{\frac{i\omega}{2}\lambda_{A} {u^B}_{\dot{A}}\lambda_{B}x^{A\dot{A}}}
\label{da3}~,
\end{align}
where the integral appearing here is precisely that which generates
${\bf X}_{ABCD}$. Using eq.~(\ref{DC2}) we then immediately find
\begin{equation}
  {\mathbf{\Phi}}_{AB\Cd\Dd}={u^C}_{\Cd}{u^D}_{\Dd}
  \left(\frac{\Phi_{(AB}\Phi_{CD)}}{S}\right).
  \label{DC3}
\end{equation}
This is indeed a double copy formula for the mixed-index Riemann
spinor. However, its form is different to formulae such as
eq.~(\ref{WeylDC}) that have previously arisen in the Weyl double copy
for vacuum pure gravity solutions. In particular, the numerator does
not simply contain a product of electromagnetic field strength
tensors, but instead involves an additional projector, that depends on
the 4-velocity of the static source. Elucidating this structure
crucially depended on transforming the momentum-space amplitude for
the JNW solution to an intermediate twistor space, as it is this that
allows us to ascertain the relation eq.~(\ref{lambdaprop}), that is
ultimately responsible for the additional prefactors in
eq.~(\ref{DC3}). Indeed, eq.~(\ref{DC3}) bears a resemblance to the
consistency relation obtained for the Coulomb solution in
eq.~(\ref{Phirel}), and this begs the question of whether
eq.~(\ref{DC3}) is fully general, or whether such a form is highly
specialised to the particular JNW solution we are considering here. We
will address this point more fully in the following section, but first
note that it is interesting to compare our results with the recent
refs.~\cite{Easson:2021asd,Easson:2022zoh}, which looked at
generalising the Weyl double copy in pure gravity to non-vacuum
solutions. Various formulae were proposed for the Riemann spinor, for
different types of solution. They included combinations such as
\begin{equation}
  \Phi_{AB\Cd\Dd}\propto \Phi_{AB}\tilde{\Phi}_{\Cd\Dd}
  \label{Manton}
\end{equation}
i.e. involving products of the two distinct gauge theory field
strength spinors. Interestingly, the formula of eq.~(\ref{DC3}) does
not have this form. To see this, note that the symmetrised brackets
can be expanded to give
\begin{equation}
    {\mathbf{\Phi}}_{AB\Cd\Dd}={u^C}_{\Cd}{u^D}_{\Dd}\frac{1}{S}
    \left(\Phi_{AB}\Phi_{CD}
    +\Phi_{AC}\Phi_{BD}+\Phi_{AD}\Phi_{BC}
    \right).
\label{DC3b}
\end{equation}
In the first term, the consistency relation of eq.~(\ref{Phirel}) may
be used to express the mixed-index Riemann spinor in terms of a
product of gauge theory field-strengths. However, this is not true for
the second and third terms, which will be non-zero in general. To see
this, one may set
\begin{displaymath}
  \Phi_{AB}\propto\alpha_{(A}\beta_{B)},\quad
  I_{ABCD}={u^C}_{\Cd}{u^D}_{\Dd}\Big[\Phi_{AC}\Phi_{BD}
    +\Phi_{AD}\Phi_{BC}\Big],
\end{displaymath}
and perform an explicit calculation to obtain
\begin{equation}
  I_{0011}\propto \alpha_0^2\beta_0^2.
  \label{I0011}
\end{equation}
If $I_{ABCD}$ is to vanish, then we must have $\alpha_0=0$ or
$\beta_0=0$, where we may choose the former without loss of
generality. We then find
\begin{equation}
  I_{0000}\propto \alpha_1^2\beta_0^2.
  \label{I0000}
\end{equation}
We cannot now choose $\alpha_1=0$ without the entire spinor field
$\mathbf{\Phi}_{AB\Cd\Dd}$ vanishing. Thus, we must choose $\beta_0=0$, which
further implies
\begin{equation}
  I_{1111}=\alpha_2^2\beta_2^2.
  \label{I1111}
\end{equation}
We are now forced to choose either $\alpha_2=0$ or $\beta_2=0$, such
that the only way that $I_{ABCD}$ can vanish is if
$\mathbf{\Phi}_{AB\Cd\Dd=0}$. Thus, the second and third terms in
eq.~(\ref{DC3b}) are indeed non-zero in general. As a consequence, the
double copy formula that the twistor methods arrive at is distinctly
different to other double-copy ans\"{a}tze appearing in the
literature. This is in itself not surprising, given the findings of
ref.~\cite{Monteiro:2021ztt}, namely that a pure double copy of gauge
theory field strengths in position space was not possible for the JNW
solution in the tensorial language.

Finally, we note that the counterpart of eq.~(\ref{DC3}) for the other
mixed-index Riemann spinor is derived to be
\begin{equation}
  \tilde{{\mathbf{\Phi}}}_{\Ad\Bd CD}={u_C}^{\Cd}{u_D}^{\Dd}
  \left(\frac{\tilde{\Phi}_{(\Ad\Bd}\tilde{\Phi}_{\Cd\Dd)}}
       {\tilde{S}}\right),
  \label{DC4}
\end{equation}
which is simply related to eq.~(\ref{DC3}), such that similar
observations to the above apply.

In this section, we have examined the JNW solution in ${\cal N}=0$
supergravity, finding that it is indeed possible to construct
spinorial double copy formulae for all spinors entering the
decomposition of eq.~(\ref{eq:curlyRspinor}). Whilst two of these
formulae are straightforward counterparts of the Weyl double copy for
vacuum solutions in pure gravity, the mixed-index double copy formulae
are different to anything encountered before. In order to probe how
general such formulae are, we must move away from the spinless and
magnetic-chargeless case of eq.~(\ref{photonamp}). This is the subject
of the following section.

\section{The case of non-zero spin and NUT charge}
\label{sec:spin}

We can generalise the Coulomb solution to include a non-zero magnetic
monopole charge and spin by using the full amplitudes of
eq.~(\ref{photonamp}), in which the parameters
$\alpha\equiv(a,\theta)$ are turned on in each theory. In order
to examine the implications of this for generalising the spinorial
double copy, it is instructive to first review the results of
ref.~\cite{Luna:2022dxo}, in pure gravity.

\subsection{The Kerr-Taub-NUT solution in pure gravity}
\label{sec:KTNgrav}

In section~\ref{sec:JNW}, we have expressed gravity amplitudes leading
to the JNW solution in terms of gauge theory amplitudes according to
eq.~(\ref{eq:DCmap}). However, for amplitudes in pure gravity
corresponding to non-zero spin and / or NUT charge, it is conventional
to write these in the form (see
e.g. refs.~\cite{Monteiro:2021ztt,Luna:2022dxo})
\begin{equation}
  {\cal M}_{\pm}\sim\frac{{\cal A}_\pm {\cal A}_{\pm}}{{\cal A}^{\rm
      scal.}_{\pm}}.
  \label{Mpuregrav}
\end{equation}
Here ${\cal A}^{\rm scal.}_{\pm}$ is a three-point amplitude for
emission of a massless scalar from a spinning and / or
``magnetically'' charged particle. The classical solution
corresponding to this amplitude is the so-called zeroth copy of the
$\sqrt{\rm Kerr}$ or Taub-NUT solution, that comprises a solution of
linearised biadjoint scalar field theory. The zeroth copy field is
indeed different for the (anti-)self-dual cases, hence the $\pm$ label
on the scalar amplitude. To understand further the reason why the
scalar amplitude is needed in eq.~(\ref{Mpuregrav}), it is sufficient
to consider the Kerr solution with spin $a^a$, for which the relevant
3-point amplitude for the anti-self-part is given by
\begin{equation}
  {\cal M}_+=e^{ik\cdot a}{\cal M}_+^{(0)},
  \label{MKerr}
\end{equation}
where ${\cal M}^{(0)}$ denotes the spinless (Schwarzschild)
amplitude. The photon amplitude for the corresponding $\sqrt{\rm
  Kerr}$ solution is
\begin{equation}
  {\cal A}_+=e^{ik\cdot a}{\cal A}_+^{(0)},
  \label{AKerr}
\end{equation}
where ${\cal A}^{(0)}_+$ denotes the Coulomb amplitude. Thus, if we
were simply to square eq.~(\ref{AKerr}) according to
eq.~(\ref{eq:DCmap}), we would instead generate the gravitational
amplitude
\begin{displaymath}
  e^{2ika}{\cal M}_+^{(0)},
\end{displaymath}
which has twice the spin of the usual Kerr solution. In order to
formulate the double copy between the conventional $\sqrt{\rm Kerr}$
and Kerr solutions, one must then introduce the scalar amplitude
\begin{equation}
  {\cal A}_+^{\rm scal.}=e^{ik\cdot a}{\cal A}^{(0),\rm scal.},
  \label{AscalKerr}
\end{equation}
where ${\cal A}^{\rm scal.}$ is the amplitude for emission of a scalar
from a spinless particle (and indeed is just a constant). The
combination of eq.~(\ref{Mpuregrav}) then performs the relevant double
copy. Reference~\cite{Monteiro:2021ztt} has argued that this leads to
an ambiguity in the double copy, namely that one may choose {\it
  different} scalar functions, such that the spin of the Kerr solution
is apportioned in different ways in the two gauge theory
amplitudes. It remains, true, however, that only one such choice
matches the original Kerr-Schild and Weyl double copies for the Kerr
solution~\cite{Monteiro:2014cda,Luna:2018dpt}. Indeed, the requirement
of having a local double copy in position space itself fixes the
relevant scalar amplitude, as follows from the arguments of
ref.~\cite{Luna:2022dxo}. There, it was observed that
eqs.~(\ref{neggraviton}, \ref{negphoton}), and their counterpart for a
massless scalar field:
\begin{equation}
  \phi=\mathrm{Re} \,i \int d\Phi(k)\hat{\delta}(2p\cdot k)
       {\cal A}^{\rm scal.}_+ e^{-ik\cdot x},
       \label{KMOCscal}
\end{equation}
generate cohomology representatives in twistor space for each theory
that are related in certain circumstances by the simple product-like
relationship
\begin{equation}
  \mathfrak{M}_{\rm grav.}=\frac{\mathfrak{M}_{\rm EM}\mathfrak{M}_{\rm EM}}
  {\mathfrak{M}_{\rm scal.}}.
  \label{twistorDC}
\end{equation}
This is precisely the twistor double copy of
refs.~\cite{White:2020sfn,Chacon:2021wbr}, which it was shown leads to
the position-space Weyl double copy of ref.~\cite{Luna:2018dpt} as a
consequence of the Penrose transform. Furthermore, the fact that each
twistor quantity $\mathfrak{M}$ arises from a precise integral
transform of a scattering amplitude provides a rule for picking out a
particular cohomology representative, thus resolving the puzzle for
how one is allowed to ``multiply'' together functions in twistor
space. 

As ref.~\cite{Luna:2022dxo} makes clear, the fact that the double copy
has a local product structure in twistor space is not generic, but
relies on the special properties of certain three-point scattering
amplitudes. We see, for example, in eq.~(\ref{Mgothic}) that the
integral that takes one from momentum space to twistor space is a
Laplace transform in the energy $\omega$. Given that the double copy
of three-point amplitudes is indeed a product in momentum-space, it
can only be true that a local product is obtained in twistor space
provided the integrands of the $\omega$ integral for each amplitude
consist of functions whose convolution is equivalent to a product of
similar functions. This is true for pure power-like functions of
$\omega$, which indeed correspond to the three-point amplitudes for
spinless particles (i.e. those leading to the Coulomb and Schwarzschild
solutions). This is not the only possibility: one is also free to
shift the conjugate variable in the Laplace transform:
\begin{equation}
  e^{\omega U}\rightarrow e^{\omega(U+V)},
  \label{Ushift}
\end{equation}
given that this operation commutes with the convolution. This is in
fact what eqs.~(\ref{MKerr}, \ref{AKerr}, \ref{AscalKerr}) do, as the
4-momentum $k^a$ is linear in $\omega$. Thus, the Newman-Janis shift
acting on a 3-point amplitude is such that it preserves the local
double copy in twistor space. Crucially, however, this will only work
if the shifts acting on the gravity, gauge and scalar amplitudes {\it
  are the same}, and the physics of this is that the corresponding
spacetime fields must be related by the conventional single and zeroth
copies. Once a local product in twistor space is obtained, a local
spacetime spinorial double copy will emerge automatically from the
Penrose transform, as shown in ref.~\cite{Luna:2022dxo}.

\subsection{Generalisation to ${\cal N}=0$ supergravity}
\label{sec:N=0Kerr}

Returning to the full spectrum of ${\cal N}=0$ supergravity, we now
wish to see whether the double copy formulas of eqs.~(\ref{DC1},
\ref{DC2}, \ref{DC3}, \ref{DC4}) generalise to the presence of
non-zero spin and / or NUT charge. Given that the arguments are
similar in both cases, we will here restrict ourselves to a non-zero
spin vector $a^a$ for each source particle. Then the KMOC formulae of
eqs.~(\ref{posgraviton}--\ref{conda}) become
\begin{align}
{\mathbf{X}}_{ABCD}
&=
-\frac{\kappa^2c_{++}}{2Q^2}\,\Re\,i\int\d \Phi(k)  \del(2 p \cdot k) \,
	e^{-ik\cdot(a_L+a_R)}\ampA_+^{(0)}\ampA_+^{(0)}\,
\ket{k}_A\ket{k}_B
\ket{k}_C\ket{k}_D\,e^{-ik\cdot x}~ \label{posgraviton2},
\\[1em]
\tilde{\mathbf{X}}_{\Ad\Bd\Cd\Dd}
&=
-\frac{\kappa^2c_{--}}{2Q^2}\,\Re\,i\,\int\d \Phi(k) \del(2 p \cdot k) 
 \,
	e^{ik\cdot(a_L+a_R)}\ampA_-^{(0)}\ampA_-^{(0)}\,
[k|_{\Ad} [k|_{\Bd}
[k|_{\Cd}[k|_{\Dd}\,e^{-ik\cdot x}~ \label{neggraviton2},
\\[1em]
{\mathbf{\Phi}}_{AB\Cd\Dd}
&=
+\frac{\kappa^2c_{+-}}{2\,Q^2}\,\Re i\int\d\Phi(k) \del(2 p \cdot k) 
 \,
	e^{-ik\cdot(a_L-a_R)}\ampA_+^{(0)}\ampA_-^{(0)}\,
\,\ket{k}_A\ket{k}_B[k|_{\Cd}[k|_{\Dd}\,e^{-ik\cdot x} \label{da4}~,
\\[1em]
{\tilde{ \mathbf{\Phi}}}_{\Ad\Bd CD}
&=
+\frac{\kappa^2c_{-+}}{2\,Q^2}\,\Re i\int\d\Phi(k) \del(2 p \cdot k)  \,
	e^{ik\cdot(a_L-a_R)}\ampA_-^{(0)}\ampA_+^{(0)}\,
\,[k|_{\Ad}[k|_{\Bd} \ket{k}_C\ket{k}_D\,e^{-ik\cdot x}~\label{conda4}, 
\end{align} 
where $a^a_{L,R}$ are the spin vectors of the two gauge theory
solutions, and ${\cal A}_{\pm}^{(0)}$ the photon amplitudes in the
spinless case. Following arguments exactly analogous to the previous
section, we can write each product of amplitudes (and spin factor) as
a combination of spinless scalar and gauge theory amplitudes, each
shifted by a common exponential factor:
\begin{equation}
  e^{\alpha_{\eta_L\eta_R}}
  {\cal A}_{\eta_L}{\cal A}_{\eta_R}\rightarrow
  \frac{[e^{\alpha_{\eta_L\eta_R}} {\cal A}^{(0)}_{\eta_L}]
    [e^{\alpha_{\eta_L\eta_R}} {\cal A}^{(0)}_{\eta_R}]
  }{[e^{\alpha_{\eta_L\eta_R}}{\cal A}^{(0),\rm scal.}]},
  \label{ALRprod}
\end{equation}
where we have defined
\begin{equation}
  \alpha_{\eta_L\eta_R}=-ik\cdot(\eta_{L}a_L+\eta_Ra_R).
  \label{alphaLRdef}
\end{equation}
It is now straightforward to apply the arguments of
section~\ref{sec:JNW}, and the result is that eqs.~(\ref{DC1},
\ref{DC2}, \ref{DC3}, \ref{DC4}) are replaced by
\begin{align}
  \tilde{\bf X}_{\Ad\Bd\Cd\Dd}[a_R+a_L]&=\frac{\Phi_{(\Ad\Bd}[a_R+a_L]
    \Phi_{\Cd\Dd)}[a_R+a_L]}{\tilde{S}[a_R+a_L]},\label{DC1b}\\
  {\bf X}_{ABCD}[a_R+a_L]&=\frac{\Phi_{(AB}[a_R+a_L]
    \Phi_{\Cd\Dd)}[a_R+a_L]}{S[a_R+a_L]},
  \label{DC2b}
\end{align}
and 
\begin{align}
  {\bf \Phi}_{AB\Cd\Dd}[a_R-a_L]&={u^C}_{\Cd}{u^D}_{\Dd}\left(
\frac{\Phi_{(AB}[a_R-a_L]\Phi_{CD)}[a_R-a_L]}{S[a_R-a_L]}
\right),\label{DC3c}\\
  {\bf \Phi}_{\Ad\Bd CD}[a_R-a_L]&={u_C}^{\Cd}{u_D}^{\Dd}\left(
  \frac{\tilde{\Phi}_{(\Ad\Bd}[a_R-a_L]\tilde{\Phi}_{\Cd\Dd)}
    [a_R-a_L]}{\tilde{S}[a_R-a_L]}
\right).\label{DC4c}
\end{align}
Here all electromagnetic spinors and scalar fields correspond to the
$\sqrt{\rm Kerr}$ and zeroth copy Kerr solutions respectively, but
where the spin is taken to be a particular combination of $a_R$ and
$a_L$, as indicated by the arguments of each field. As for the JNW
solution considered in the previous section, the double copy formula
for the mixed-index spinors involves a pre-factor that depends upon
the 4-velocity of the source particle. Interestingly, this is the same
factor that appears in the spinless case, a fact which is ultimately
due to the static nature of the solution, in that the prefactor arises
from the delta function $\delta(2p\cdot k)$ in
eqs.~(\ref{posgraviton2}--\ref{conda4}). However, the spin arguments
in eqs.~(\ref{DC1b}--\ref{DC4c}) are such that these formulae do not
admit a strict double-copy interpretation. In order to satisfy the
requirements of a local position-space double copy, as outlined in the
previous section, the spin parameters $a^a$ of each electromagnetic
spinor and scalar field must be the same. This in turn means that the
combinations of spin vectors appearing in each individual
electromagnetic spinor depend on {\it both} $a_R$ and $a_L$ i.e. the
spin vectors from both the gauge theories appearing in the double
copy. There is, of course, a special case in which we may indeed
obtain a double-copy interpretation, namely $a_R=a_L$. This matches
what happens for the Kerr solution in pure gravity, but is such that
the axion will automatically vanish. Furthermore, the dilaton will
look like a dilaton that is generated by a non-spinning particle, and
it is not clear if such a solution can be made to satisfy the
Einstein-dilaton equations of motion beyond linearised order (see
e.g. ref.~\cite{Bogush:2020lkp} for related work).

It is perhaps worth stressing that eqs.~(\ref{DC1b}--\ref{DC4c}), even
if they lack a strict double copy interpretation, nevertheless
constitute exact position-space relations for solutions of ${\cal
  N}=0$ supergravity at linearised order. They may therefore be useful
for something, however restricted in scope.

\section{Discussion}
\label{sec:discuss}

In this paper, we have addressed the question of whether
position-space double copy formulae exist for ${\cal N}=0$
supergravity, that are analogous to the well-known Weyl double
copy~\cite{Luna:2018dpt}. To this end, we have used recently developed
methods that express classical solutions in terms of on-shell inverse
Fourier transforms of scattering amplitudes~\cite{Kosower:2018adc},
together with arguments that split this transform into two
stages~\cite{Guevara:2021yud}. The first stage takes amplitudes into
twistor space, such that the twistor double copy of
refs.~\cite{White:2020sfn,Chacon:2021wbr} is obtained. The second
stage then consists of the well-known Penrose transform from twistor
to position space, and allows one to discern a position-space double
copy formula, if it exists. This chain of arguments has been
previously used to derive the original Weyl double
copy~\cite{Luna:2022dxo}, and also the so-called Cotton double copy
for topologically massive solutions in three spacetime
dimensions~\cite{Emond:2022uaf,Gonzalez:2022otg,CarrilloGonzalez:2022ggn}. Thus,
it is natural to try to apply it to the case of ${\cal N}=0$
supergravity, which after all is the known ``full'' double copy of
Yang-Mills theory.

Whether or not a position-space double copy exists for ${\cal N}=0$
supergravity has been recently considered in
ref.~\cite{Monteiro:2021ztt}, which indeed inspired the present
study. The authors in that case found that no double copy exists if
the dilaton and / or axion are turned on, where the tensorial
formalism was used. Our arguments in this paper show that this is not
quite true, and that one can indeed write double copy formulae for all
spinor fields appearing in the generalised Riemann tensor of
eq.~(\ref{eq:curlyRspinor}), at least for the JNW solution sourced by
a spinless particle with no NUT charge. For those spinors with a
single type of index, the formulae precisely mirror those for the case
of pure gravity. For the mixed-index spinors, however, there are
additional factors involving the 4-velocity of the source
particle. That these were not considered in
ref.~\cite{Monteiro:2021ztt} may be due to its focus on tensorial
formulae for the position-space double copy, given that the
translation from spinors to tensors can obscure simple properties in
the former language. Furthermore, the twistor methods considered here
proved crucial in deriving the presence of the additional prefactor,
which we note is also absent in recent conjectures for how to
double-copy spinors for non-vacuum gravity
solutions~\cite{Easson:2021asd,Easson:2022zoh}.

When non-zero spin and / or NUT charge are present, it is yet again
possible to write formal position-space double copy formulae for
solutions of ${\cal N}=0$ supergravity, which again involve similar
prefactors to the JNW case. However, the interpretation of these
formulae is not natural, given that they must involve products of
electromagnetic spinors, each of which involves the spin / NUT
parameters of the full gravity solution. Whether or not such formulae
are useful is a matter of debate, but it is in any case interesting
that the twistor methods, as in
refs.~\cite{Luna:2022dxo,CarrilloGonzalez:2022ggn}, are again able to
ascertain both the presence of spinorial double copy formulae, but
also their scope and applicability. It would be interesting to
investigate whether similar methods could shed light on the
generalised (non-vacuum) double copies explored in
refs.~\cite{Easson:2021asd,Easson:2022zoh}, or indeed to other
theories and / or types of solution.


\section*{Acknowledgments}

We thank Tucker Manton and Ricardo Monteiro for helpful
discussions. This work has been supported by the UK Science and
Technology Facilities Council (STFC) Consolidated Grant ST/P000754/1
``String theory, gauge theory and duality''. KAW is supported by a
studentship from the UK Engineering and Physical Sciences Research
Council (EPSRC).

\bibliography{refs}
\end{document}